\documentclass[final,5p,twocolumn]{elsarticle}
\usepackage{epsfig}
\usepackage{amssymb}
\usepackage{amsthm}
\usepackage{lineno}
\usepackage{amsmath,graphicx,url,color}

\journal{Computers and Geosciences}

\begin{document}

\begin{frontmatter}

\title{\texttt{CoinCalc} -- A new R package for quantifying simultaneities of event series}


\author[1,2]{Jonatan F. Siegmund}\ead{jonatan.siegmund@pik-potsdam.de}
\author[2,3,4]{Nicole Siegmund}
\author[1]{Reik V. Donner}
\address[1]{Research Domain IV -- Transdisciplinary Concepts and Methods, Potsdam Institute for Climate Impact Research, Telegrafenberg A31, 14473 Potsdam, Germany}
\address[2]{Institute of Earth and Environmental Science, University of Potsdam, Karl-Liebknecht-Stra{\ss}e 24-25, 14476 Potsdam-Golm, Germany}
\address[3]{Leibniz Centre for Agricultural Landscape Research, Department for Soil Landscape Reseach, Eberswalder Stra{\ss}e 84, 15374 M\"uncheberg, Germany}
\address[4]{Institute of Meteorology and Climate Research, Atmospheric Environmental Research (IMK-IFU), Karlsruhe Institute of Technology, Kreuzeckbahnstra{\ss}e 19, 82467 Garmisch-Partenkirchen, Germany }

\begin{abstract}
We present the new R package \texttt{CoinCalc} for performing event coincidence analysis (ECA), a novel statistical method to quantify the simultaneity of events contained in two series of observations, either as simultaneous or lagged \emph{coincidences} within a user-specific temporal tolerance window. The package also provides different analytical as well as surrogate-based significance tests (valid under different assumptions about the nature of the observed event series) as well as an intuitive visualization of the identified coincidences. We demonstrate the usage of \texttt{CoinCalc} based on two typical geoscientific example problems addressing the relationship between meteorological extremes and plant phenology as well as that between soil properties and land cover. 
\end{abstract}


\end{frontmatter}

\section{Introduction}
In many areas of geosciences, but also other scientific disciplines like neurosciences, there has been a rising interest in inferring information on dynamical interdependencies between different observational series that are not given in the form of continuous or discrete-valued time series, but as sequences of events (e.g., unmarked or marked point processes). Traditional statistical tools like classical (Pearson) correlation analysis are often not directly applicable to such series or of limited explanatory value. While in neurosciences, many methodological developments have been introduced and subsequently applied for studying the statistical interrelationships between event series (e.g., describing sequences of neuronal spiking activity \citep{Brown2004,Lewicki1988}), there have been relatively few attempts to transfer corresponding approaches to geoscientific problems \citep{Boers2015,Malik2010a}.

Event coincidence analysis (ECA) is a recently developed method for studying the statistical interdependency between two event series, which has been originally introduced and applied in a geoscientific context \citep{Donges2011,Donges2015,Rammig2014,Siegmund2015}. Unlike correlation analysis, this method exclusively takes the timings of certain well-defined events in two series into account and ignores potentially available other information (e.g., underlying explicit time series values) on the gradual variability of related observables. Therefore, it provides a complementary view on data that are either by definition of binary structure (event/no event) or where only certain values (e.g., extreme events) are expected to result in a specific response of interest. Examples include the timings of natural disasters like earthquakes or floods \citep{Donges2015} or cases where strong deviations from ``normal'' behavior can result in qualitatively different interdependencies between the variables of interest (e.g., ecosystem responses to extreme environmental conditions like droughts, cold spells or volcanic eruptions) \citep{Reichstein2013,Zscheischler2013}. 

So far, ECA has been successfully applied to studying problems in biogeoscientific \citep{Rammig2014,Siegmund2015}, socio-ecological \citep{Donges2015} and paleoclimatic contexts \citep{Donges2011}. The diversity of research questions discussed in the aforementioned publications suggests a wide range of possible future applications. While \citet{Rammig2014} and \citet{Siegmund2015} used the approach to derive complementary information (beyond classical correlation analysis) by looking at the timing of extreme events, the analyses of \citet{Donges2011,Donges2015} could not have been conducted using standard tools of classical statistics since they addressed series of explicit events.

This paper introduces \texttt{CoinCalc}, an easy-to-handle implementation of ECA in the open statistical software R, which is available via the Comprehensive R Archive Network (CRAN, \url{www.r-project.org}). We emphasize that the CRAN repository already contains the package \texttt{CNA} for performing an entirely different type of analysis referred to as \emph{coincidence analysis} \citep{Baumgartner2015}, and that the same term is also used in particle physics \citep{Zaborov2009} in yet another different context. Within the framework of \texttt{CoinCalc}, we exclusively refer to the definition of \emph{event} coincidence analysis as comprehensively described by \citet{Donges2015}.

The remainder of this paper is organized as follows: In Sect.~\ref{sec:meth}, the methodological background of ECA is provided, followed by a detailed description of the functions provided by \texttt{CoinCalc} and their options in Sect.~\ref{sec:func}. Finally, two exemplary applications of the package to different geoscientific data sets are discussed in Sect.~\ref{sec:examples}. The paper concludes with a short summary in Sect.~\ref{sec:conclusions}.

\section{Methodological Background} \label{sec:meth}

\subsection{Event series with continuous and discrete event times}
Let us consider two sequences of events of distinct types $A$ and $B$ that occur at times $t_i^A$ and $t_j^B$ with $i=1,\dots,N_A$ and $j=1,\dots,N_B$, where $N_{A}$ and $N_{B}$ are the number of events of type $A$ and $B$, respectively. Here, we exclusively focus on the timing of events and disregard any possibly available information on the magnitudes of these events. 

Depending on the specific question under study, event series can be given in terms of two generic data types corresponding to either continuous or discrete timings of events. On the one hand, we may have just a list of event times (e.g., in \cite{Donges2011}) with no continuously observed data between these events. We will refer to this type of data as \emph{event sequences} in the following. On the other hand, we may have the situation of a time series containing time-discrete observations of a possibly continuously-valued variable upon which events are defined. The resulting \emph{event time series} is conveniently represented by a binary sequence of a length $T$ equal to the number of underlying observations, where entries 1 (0) correspond to time steps with (without) an event.

\subsection{Counting Coincidences}

ECA essentially counts how often events occur in both series \emph{simultaneously} (referred to as \emph{coincidences}). The notion of \emph{simultaneity} can be further specified by considering two parameters: a user-defined time lag $\tau$ and a certain tolerance window $\Delta T$. The consideration of $ \tau\neq 0$ can be important in order to study lagged responses of events of type $A$ to such of type $B$ (or vice versa) as known for, e.g., the energy exchange between hydrosphere and atmosphere \citep{Iwi2006,Kumar2003,Wedgbrow2002} or various ecological systems \citep{Boulton2003,Daan2005,Letnic2005}. In turn, $\Delta T$ allows addressing uncertain timings of events in examples like climate reconstructions \citep{Jones2009,Woodborne2015}, archaeological, paleontological or paleoanthropological records \citep{Donges2011}, or events with an extended duration like climate regime shifts and ecological or social responses to natural disasters \citep{Donges2011,Donges2015}.

By definition, the notion of event coincidence is not symmetric, i.e., always takes one of the two event series as a reference to which the second is compared. Commonly, in this context events of type $B$ are considered as possibly influencing the timings of events of type $A$, and \emph{not} vice versa (of course, the roles of both variables might be interchanged). However, there might be applications where such a presumed directional influence cannot be postulated in advance -- in this case, ECA can be utilized as an explanatory rather than confirmatory statistical tool to test for the existence, direction and significance of such relationships.

Following this conceptual idea, ECA distinguishes between the \emph{precursor coincidence rate} 

\begin{equation}\label{eq1}
 r_{p}(\Delta T,\tau)= \frac{1}{N_{A}} \sum_{i=1}^{N_{A}}\Theta \left(\sum_{j=1}^{N_{B}}1_{[0,\Delta T]}((t_{i}^{A}-\tau)-t_{j}^{B})\right)
\end{equation}
\noindent
and the \emph{trigger coincidence rate} 

\begin{equation} \label{eq2}
 r_{t}(\Delta T,\tau)= \frac{1}{N_{B}} \sum_{j=1}^{N_{B}}\Theta \left(\sum_{i=1}^{N_{A}}1_{[0,\Delta T]}((t_{i}^{A}-\tau)-t_{j}^{B})\right),
\end{equation}
\noindent
where $\Theta(\cdot)$ is the Heaviside function and $1_{[0,\Delta T]}$ is the indicator function of the interval $[0,\Delta T]$. For $\Delta T=0$, the term in the inner sum will just collapse to $\delta(t_{i}^{A}-\tau,t_{j}^{B})$, where $\delta(\cdot,\cdot)$ is the Kronecker delta, providing a value of 1 if and only if both arguments are equal, and zero otherwise. In this context, $r_{p}(\Delta T,\tau)$ denotes ``the fraction of $A$-type events that are preceded by at least one $B$-type event'', while  $r_{t}(\Delta T,\tau)$ measures ``the fraction of $B$-type events that are followed by at least one $A$-type event'' \citep{Donges2015}. By definition, both coincidence rates can only take values between 0 (complete absence of coincidences) and 1 (all events coincide with events in the reference series).

\subsection{Significance Tests}
Beyond the sole calculation of coincidence rates, \texttt{CoinCalc} currently provides three significance tests that can be selected in order to comply with the specific properties of the event series under study. 

\subsubsection{Analytical Test: Poissonian approximation}

Under the assumption that the $A$ and $B$-type events are randomly distributed \emph{and} mutually independent (i.e., follow two independent Poisson processes) and sufficiently rare, the probability of observing a given number of precursor coincidences $K_{p}=N_{A} \cdot r_{p}$ can be approximated by a binomial distribution as

\begin{equation} \label{eq3}
\begin{split}
 P(K_{p}) = & \left({N_{A}\atop K_{p}}\right)\left(1-\left(1-\frac{TOL}{T-\tau}\right)^{N_{B}}\right)^{K_{p}} \times \\
& \qquad \times \left(\left(1-\frac{TOL}{T-\tau}\right)^{N_{B}}\right)^{N_{A}-K_{p}}.
\end{split}
\end{equation}
Here, all time values are given either in absolute time units (for event sequences) or as discrete numbers of time steps (for event time series). Accordingly, we have $TOL=\Delta T$ for event sequences and $TOL=\Delta T +1$ for event time series. In a similar way, $T$ denotes either the total time span of observations (for event sequences) or the number of observations (for event time series). Note that while $TOL$ is a non-negative parameter that can be selected according to the specific problem under study, $T$ is itself part of the necessary information on the event series under study that needs to be known in order to perform ECA.

The $p$-value of the corresponding \emph{analytical} significance test provided by \texttt{CoinCalc} corresponds to the probability that $K_{p}$ \emph{or more} coincidences occur due to chance according to Eq.~(\ref{eq3}), 

\begin{equation}
p_{K_p} = \sum_{K_p'\geq K_p} P(K_p'),
\end{equation}
\noindent
where $K_p$ is the number of precursor coincidences obtained when comparing the empirically found event sequences $A$ and $B$. The $p$-value for the corresponding significance test of the trigger coincidence rate $r_t=K_t/N_B$ is obtained in the same way by interchanging $N_A$ and $N_B$ in Eq.~(\ref{eq3}) and replacing $K_p$ by $K_t$ \citep{Donges2015}. In both cases, the null hypothesis of the test is that the observed number of coincidences can be explained by two independent series of randomly distributed events. If the given $p$-value is smaller than a user-defined confidence level $\alpha$, this null hypothesis can be rejected.

\subsubsection{Surrogates with random event times: Shuffle test}

A rejection resulting from the analytical significance test described above can have two possible implications: either the two event series are not independent of each other (in most cases the desired type of information), or the analytical approximation does not hold. The latter problem appears, for example, if the number of events is too large in comparison to the full window of observations and the associated sampling interval, i.e., if the events cannot be considered rare \citep{Donges2015}. This is the case if the implicit conditions $N_A \Delta T, N_B \Delta T \ll T$ -- under which the definition of events is meaningful -- are violated.

In order to cope with data sets of the latter type, \texttt{CoinCalc} provides a second significance test based on the generation of an ensemble of surrogate event series where only the numbers of events in both series ($N_A$ and $N_B$) are prescribed and the actual event times are selected uniformly at random from the time interval of observations. By construction, in the limit $T\to\infty$, the waiting time distribution of such surrogate event series would be exponential corresponding again to a Poisson process. The empirical distribution $\hat{P}_{surr}(K_{p})$ of the resulting surrogate precursor coincidence rates from all pairs of these ``shuffled'' surrogate event series approximates the distribution of coincidence rates that would result from two event series with the same length and number of events as the original data sets where the individual events are completely independent of each other. Hence, the $p$-value for the precursor coincidence rate can be approximated as

\begin{equation}
p_{K_p}=1-\hat{P}_{surr}(K_{p}).
\end{equation}
\noindent
For trigger coincidences, $p_{K_t}$ follows in full analogy.

\subsubsection{Surrogates with prescribed waiting time distributions}

Besides the too large number of events in a series, another possible reason for the analytical significance test to provide incorrect results is that the empirical distribution of waiting times between subsequent events can show deviations from the exponential behavior expected for Poisson processes. For such cases, \texttt{CoinCalc} provides a third significance test based on another type of surrogate event series that resemble the original data sets in pertaining their series length and waiting time distributions. Specifically, each surrogate event series is produced by iteratively selecting the waiting time until the next event from the empirical distribution of waiting times of the original data sets uniformly at random. The calculation of $p$-values then follows the same strategy as described above for the shuffle surrogates.

Unlike analytical and shuffle tests, the latter surrogate-based test does not make any assumption about the Poissonian nature of the event series and is insofar more generally applicable and, hence, less restrictive. Moreover, as the shuffle test, it allows to perform ECA for event series that do not match the condition of rare events. However, it still provides only approximate results for cases in which there is evidence for correlations between the events of one of the series. This situation requires the utilization of numerical approximations of the distribution of the considered test statistics making use of even more sophisticated resampling approaches \citep{Donges2015}. A corresponding extension of the currently implemented significance tests is scheduled for future versions of \texttt{CoinCalc}.

\subsection{Symmetric Tolerance Windows}

Extending upon the methodological setting used by \citet{Donges2015}, \texttt{CoinCalc} also provides the possibility to define a symmetric tolerance window. The definition of such a window is of specific interest for analyses in which there is evidence for uncertainties in the timing of the events in one or both data sets. In Eqs. (\ref{eq1}) and (\ref{eq2}) discussed above, $\Delta T$ is supposed to define a non-symmetric window, thus either preceding or following the time step of interest, resulting in tolerance windows $[t_i^A-\tau-\Delta T,t_i^A-\tau]$ for a precursor coincidence and $[t_j^B+\tau,t_j^B+\tau+\Delta T]$ for a trigger coincidence.

In turn, for $\tau = 0$, a symmetric tolerance window corresponds to counting coincidences of events in series $B$ falling into a time window including a time interval $\Delta T$ both \emph{before and after} an event in series $A$. Here, the definition of the time window using this approach is $[t_i^A-\Delta T, t_i^A+\Delta T]$ (i.e., the symmetric tolerance windows are twice as large as their directional counterparts discussed above). For intervals centered around $t_j^B$ and for delayed coincidences with $\tau\neq 0$, the corresponding modifications are straightforward. Hence, Eq.~(\ref{eq1}) (and similarly Eq.~(\ref{eq2})) can be rewritten as

\begin{equation}\label{eq4}
 r_{p}(\Delta T,\tau)= \frac{1}{N_{A}} \sum_{i=1}^{N_{A}}\Theta \left(\sum_{j=1}^{N_{B}}1_{[-\Delta T,\Delta T]}(t_{i}^{A}-t_{j}^{B})\right).
\end{equation}
\noindent
As a consequence, for the calculation of $P(K_{p})$ in Eq.~(\ref{eq3}) with a symmetric tolerance window, we have $TOL =2\Delta T$ for event sequences and $TOL=2\Delta T +1$ for event time series.

\section{Description of the package} \label{sec:func}

The R package \texttt{CoinCalc} provides all necessary functionality to perform the calculation of coincidence rates and the associated significance tests according to the user's specific requirements. Specifically, it is possible to perform ECA for both event sequences (\texttt{es} format) and event time series (\texttt{ts} format).

To illustrate these two different data types, let us suppose a set of 15 equidistant observations, where the 4th, 6th, 7th and 11th values correspond to events. The corresponding event time series would read \texttt{(0,0,0,1,0,1,1,0,0,0,1,0,0,0,0)}, whereas the associated event sequence object would be \texttt{\{(4,6,7,11), span(1,15)\}}, i.e., an object containing a list of event times as well as a vector of length 2 with the start and end point of observations. Note that the latter vector is essential for performing the significance tests and, hence, meaningfully interpreting the obtained coincidence rates.

Notably, the consideration of the \texttt{es} format is particularly useful for large data sets, since the resulting computational demands are considerably lower than for \texttt{ts} data. In turn, the disadvantage of the \texttt{es} format is that in the current implementation of \texttt{CoinCalc}, the data set must be based on continuous observations with no missing periods of recording. In the case of missing observations, the \texttt{ts} format should be used instead. In general, performing ECA between two data sets requires the corresponding data being given in the same format. For this purpose, \texttt{CoinCalc} provides functions for transforming each of the two formats into the other.

In the following, we give a brief overview on the functions currently provided by the package as well as their usage and possible options: \\

\noindent
\texttt{CC.binarize()}: This function binarizes a numerical vector (i.e., a time series of an arbitrarily distributed variable) using a given threshold. This threshold can either be a percentile of the variable's empirical distribution or a specific prescribed value. The output object has \texttt{ts} format (i.e., is a binary vector of the same length $T$ as the original data). The following arguments and options need to be provided:
\begin{itemize}
\item \texttt{data}: Numerical vector (time series) to be binarized.
\item \texttt{ev.def}: String specifying the event definition method. If \texttt{ev.def="percentile"} (default), events are defined using the value of \texttt{thres} (see below) as percentile threshold. If \texttt{ev.def="absolute"}, events are defined according to an absolute threshold value \texttt{thres}.
\item \texttt{thres}: Binarization threshold. If \texttt{ev.def="percentile"}, \texttt{thres} must be a real number within [0,1]. For \texttt{ev.def="absolute"}, it can take any real number within the range of \texttt{data}.
\item \texttt{event}: String specifying whether values \texttt{"higher"} (default) or \texttt{"lower"} than \texttt{thres} are to be considered as events.
\end{itemize}

\noindent
\texttt{CC.ts2es()}: This function converts an event time series (\texttt{ts} format) into an event sequence (\texttt{es} format). It has only a single argument:
\begin{itemize}
\item \texttt{data}: Binary event time series to be transformed into an event sequence.
\end{itemize}

\noindent
\texttt{CC.es2ts()}: This function converts an event sequence (\texttt{es} format) object into a binary event time series (\texttt{ts} format). It requires the following arguments:
\begin{itemize}
\item \texttt{data}: Event sequence comprising event positions to be transformed into an event time series.  
\item \texttt{span}: Numerical vector with two elements (\texttt{span[1]}: starting point of the data set, \texttt{span[2]}: end point of data set).
\item \texttt{es.round}: Number of digits for rounding the given values in \texttt{data}. \texttt{es.round} additionally defines the temporal resolution (e.g., \texttt{es.round=3} leads to a sampling interval of 0.001 time units).
\end{itemize}

\noindent
\texttt{CC.eca.es()}: This function performs the actual ECA using two event sequences (\texttt{es} format). The arguments and options are listed below:
\begin{itemize}
\item \texttt{seriesA}: Numerical vector specifying the timings of events of type $A$.
\item \texttt{seriesB}: Numerical vector specifying the timings of events of type $B$.
\item \texttt{spanA}: Numerical vector of length 2 specifying the start and end points of the data set given in \texttt{seriesA}.
\item \texttt{spanB}: Numerical vector of length 2 specifying the start and end points of the data set given in \texttt{seriesB}.
\item \texttt{delT}: Non-negative real number for defining the tolerance window $\Delta T$. If \texttt{delT=0} (default), only simultaneous coincidences are counted. 
\item \texttt{sym}: Boolean variable (default: \texttt{FALSE}) specifying if the temporal tolerance window should be taken symmetrically or not. 
\item \texttt{tau}: Non-negative real number (default: \texttt{0}) specifying the time lag $\tau$.
\item \texttt{sigtest}: String specifying the type of significance test. If \texttt{sigtest="poisson"} (default), the analytical significance test based on the assumption of independent and sparse Poisson processes is performed. If \texttt{sigtest="shuffle"}, the test statistics for randomly located events is numerically approximated. If \texttt{sigtest="surrogate"}, the numerical approximation of the test statistics based on coincidence rates for an ensemble of surrogate event sequences with the same waiting time distributions as the original data is utilized.
\item \texttt{reps}: Positive integer (default: 1000) specifying the surrogate ensemble size for the numerical significance test.
\item \texttt{alpha}: Desired confidence level (default: $\alpha$=0.05) for the significance test specified by \texttt{sigtest}.
\end{itemize}

\noindent
\texttt{CC.eca.ts()}: This function performs ECA using two event time series (\texttt{ts} format). Missing values are allowed. If NAs are found in one of the input time series, the corresponding time steps (as well as their respective counterparts in the second series) are ignored in the performed analysis. The following arguments and options are required:
\begin{itemize}
\item \texttt{seriesA}: Binary vector specifying steps with and without events in series $A$.
\item \texttt{seriesB}: Binary vector specifying steps with and without events in series $B$.
\item \texttt{delT}:  Non-negative integer (default: 0) for defining the tolerance window $\Delta T$ (number of time steps accepted as time difference). For \texttt{delT=0}, only simultaneous events are counted as coincidences.
\item \texttt{sym}: see \texttt{CC.eca.es()}
\item \texttt{tau}:  Non-negative integer (default: \texttt{0}) specifying the time lag $\tau$.
\item \texttt{sigtest}:  see \texttt{CC.eca.es()}
\item \texttt{reps}: see \texttt{CC.eca.es()}
\item \texttt{alpha}: see \texttt{CC.eca.es()}
\end{itemize}

\noindent
\texttt{CC.plot()}: This function creates a visualization of the event series and the results of ECA. The output graphics displays which individual events correspond to coincidences. \texttt{CC.plot()} is currently only available for data given as event time series (\texttt{ts} format); an extension to general event sequences is planned for a future version of \texttt{CoinCalc}. The arguments and options are listed below:
\begin{itemize}
\item \texttt{seriesA}: see \texttt{CC.eca.ts()}
\item \texttt{seriesB}: see \texttt{CC.eca.ts()}
\item \texttt{delT}: see \texttt{CC.eca.ts()}
\item \texttt{sym}: see \texttt{CC.eca.ts()}
\item \texttt{tau}: see \texttt{CC.eca.ts()}
\item \texttt{dates}: Vector of length $T$, containing characters or numerical values (default = \texttt{NA}) providing date information for the two series. If specified, event dates are added to the plot.
\end{itemize}
\noindent
One example of an illustrative visualization produced by \texttt{CC.plot()} can be found in Sect.~\ref{sec:examples}

\section{Examples} \label{sec:examples}

\subsection{Plant phenology and meteorological extremes}\label{example2}

As a first example for the application of the \texttt{CoinCalc} package, let us reconsider the problem studied by \citet{Siegmund2015}, where ECA was used to identify time windows during the year within which unusually warm (cold) weather conditions can result in (i.e., coincide with) very early (late) flowering of central European shrub species in the same year. The information on flowering dates was provided by the German Weather Service (DWD) \citep{DWD2009}, and the temperature data were obtained by an area-weighted interpolation of temperature data from DWD-operated meteorological stations \citep{Oesterle2006}. 

\begin{figure}
 \centerline{\includegraphics[width=0.95\columnwidth]{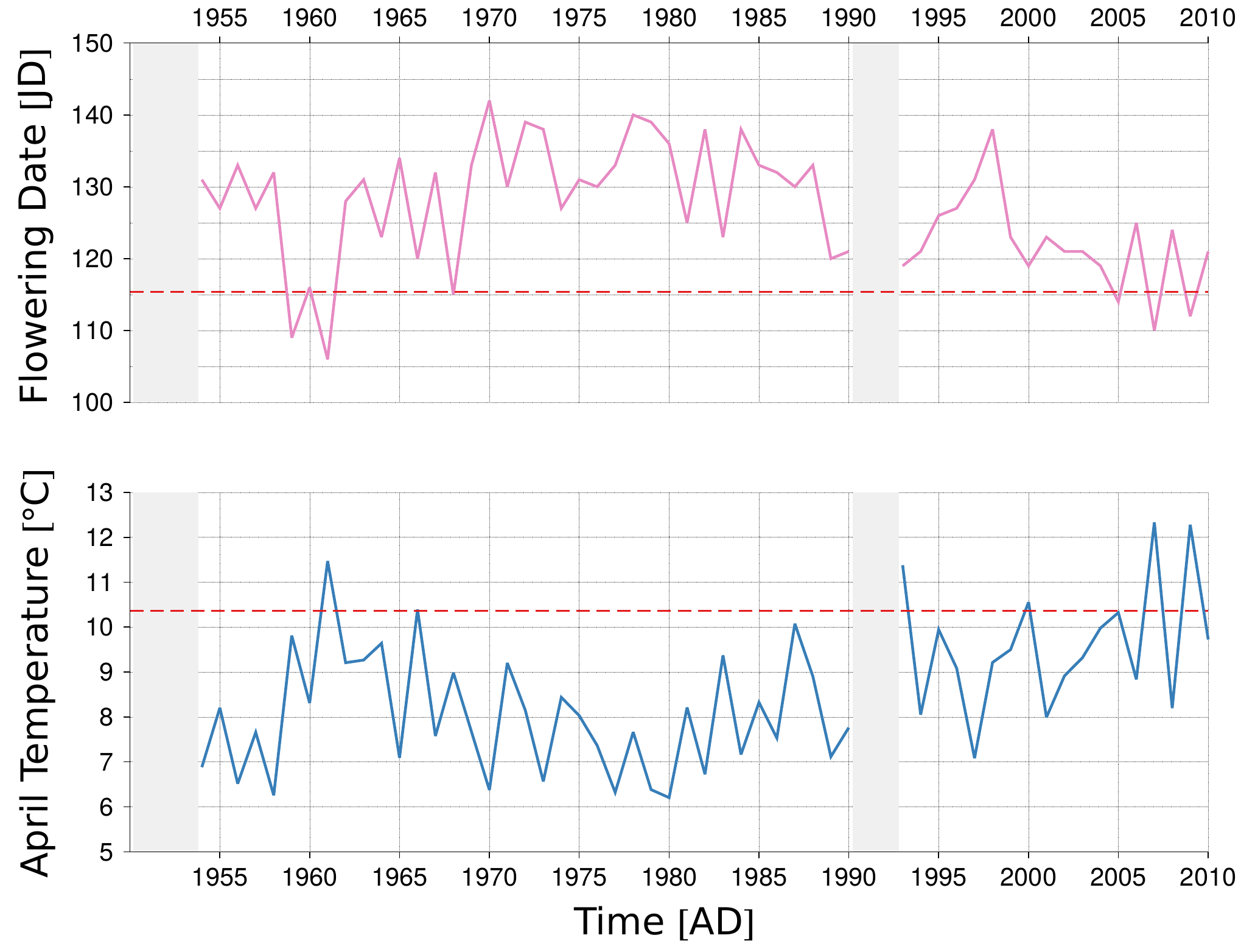}}
  \caption{Time series of Lilac flowering (Julian Day (JD) of the year, upper panel) and April mean temperature (lower panel) in Niederrimbach from 1951 to 2010. The red dashed lines mark the thresholds at the empirical 10th and 90th percentiles, respectively. Events in both time series are defined as those values, that are lower (higher) than the respective threshold.}\label{fig3}
\end{figure}

Here, we illustrate the utilization of \texttt{CoinCalc} taking just one time series of annual Lilac flowering dates and one April mean temperature time series from 1950 to 2010 as an example. Specifically, we use the data from a phenological station located in Niederrimbach, Germany (49.4833$^{\circ}$N, 10.000$^{\circ}$E). Events in the two considered time series correspond to ``very early flowering'' and ``very warm conditions'', respectively. While the former is defined as a flowering occurring earlier in the year than the empirical 10th percentile of all historical flowering dates in this record, a mean April temperature is considered to be very warm if it exceeds the 90th percentile of all observed April values. Figure \ref{fig3} shows the two time series as well as the two thresholds for the definition of events.

In order to conduct ECA for these two time series using \texttt{CoinCalc}, let the flowering data be stored in the vector \texttt{Fl.60y} and the temperature data in the vector \texttt{TT.60y}. Since a lagged effect of spring temperatures on flowering time in a yearly resolved data set is not expected, $\tau$ and $\Delta T$ take their default values of 0. Thus, executing 
\begin{verbatim}
Fl.60y.bin <- CC.binarize(data=Fl.60y, ...
... ev.def="percentile", ...
... thres=0.10, event="lower")
TT.60y.bin <- CC.binarize(data=TT.60y, ...
... ev.def="percentile", ...
... thres=0.90, event="higher")
ca.out <- CC.eca.ts(Fl.60y.bin, TT.60y.bin,...
	...sigtest="poisson")
\end{verbatim}
\noindent
results in the list \texttt{ca.out} containing the following information:
\begin{verbatim}
NH precursor: FALSE
NH trigger: FALSE
p-value precursor: 0.01777557
p-value trigger: 0.01777557
precursor coincidence rate: 0.5
trigger coincidence rate: 0.5
\end{verbatim}

In this example, the null hypotheses of independent random event series can be rejected for both precursor and trigger coincidence rates at a confidence level of $\alpha=0.05$. In the specific setting considered here, the two coincidence rates and their resulting $p$-values are identical, since both time series have the same length, no tolerance window is considered ($\Delta T = 0$) and the same number of events $N_A=N_B$ is present in both series due to their definition using the empirical 10th and 90th percentile, respectively. 

It may be worth noting that in the present example, the correlation coefficient between both original time series is already $-0.83$, suggesting the existence of a strong correlation between the two considered variables. However, such a strong correlation does not necessarily imply the co-occurrence of extreme values in both records, since correlations take all parts of the distribution into account. Hence, even with a generally strong correlation, certain parts of the distribution of the two observables can still completely mismatch in terms of their appearance in the series. Thus, ECA is a prospective complementary tool providing information to understand the relationship between distinct parts of the distribution of two time series. 

\begin{figure*}
 \centerline{\includegraphics[width=0.95\textwidth]{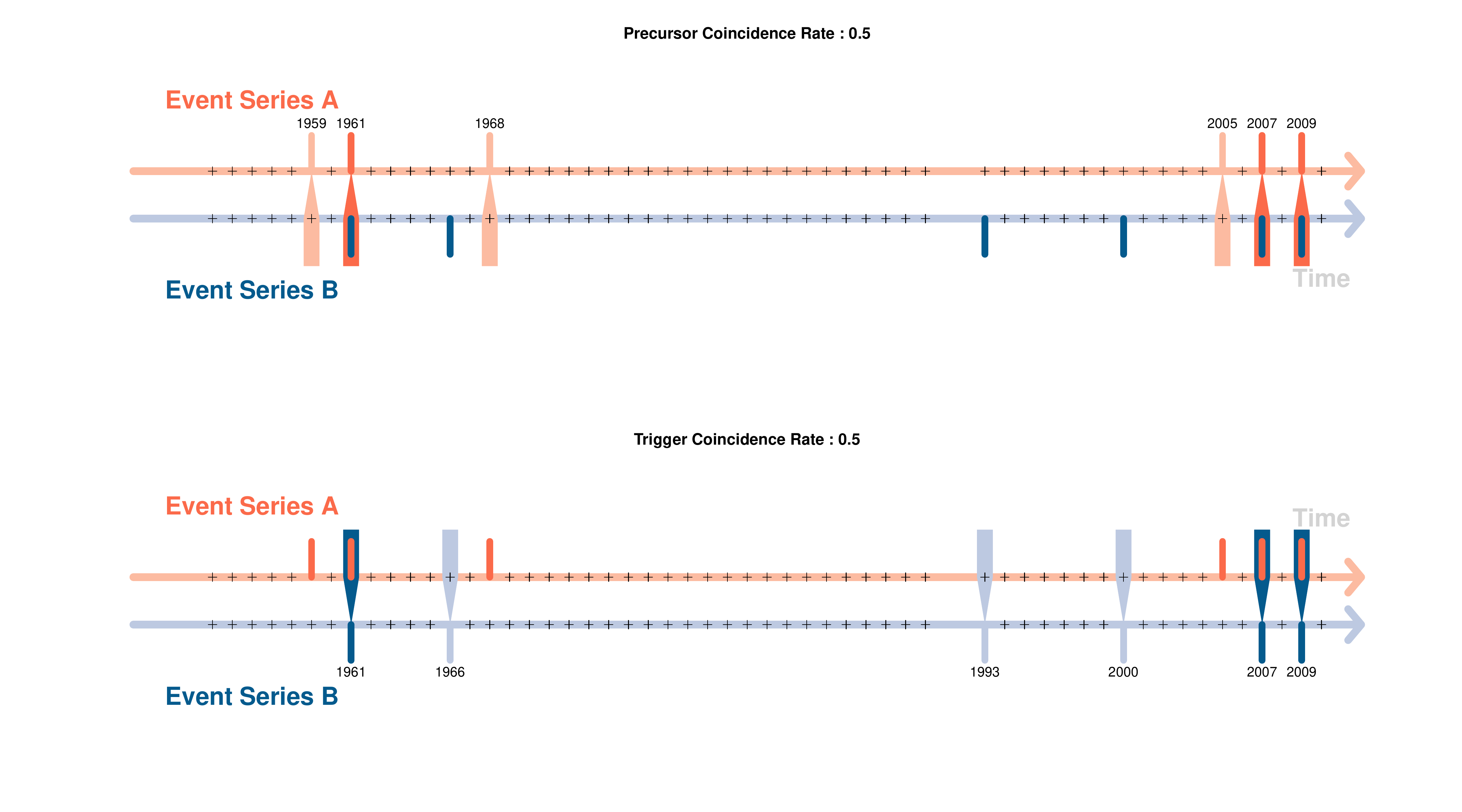}}
  \caption{Graphical output of the \texttt{CC.plot} function for the flowering (red) and temperature time series (blue). Blue and red bars mark time steps with events. For the series to be compared with the respective reference sequence, light colors show events without coincidence and dark colors coincidences. For the reference series, no corresponding visual distinction is made. Note that in the present example, $\Delta T=0$ and $\tau = 0$, i.e., only simultaneous events (referring here to the same year) are considered as coincident.}\label{fig4}
\end{figure*}

Figure \ref{fig4} shows the plot generated by the function call
\begin{verbatim}
CC.plot(Fl_60y.bin,TT.60y.bin,dates=FL.dates)
\end{verbatim}
\noindent
where \texttt{FL.dates} is a vector containing a sequence from 1951 to 2010. The light blue and light red bars indicate events in the phenological ($A$) and temperature ($B$) time series, respectively, where dark colors highlight those events that correspond to coincidences. It can be seen that in the present example, three of the six early flowering events (series $A$) coincide with warm April temperatures (series $B$) (and vice versa), yielding precursor and trigger coincidence rates of 0.5 as already given above.

\subsection{Soil organic carbon content and land-use}

Our second example illustrates the application of \texttt{CoinCalc} to a quite different problem. Here, ECA is not used for studying two time series, but a set of soil samples providing information about the soil organic carbon (SOC) contents of 218 smallhold farmer sites in the Nyando district, Western Kenya. The samples were collected, processed and analyzed in 2013/2014 in the course of the project SAMPLES (\url{http://samples.ccafs.cgiar.org}). In addition to the carbon data set, information on the crop types planted on the sampled plot was collected. During field work, two hypotheses were drawn: (i) The plantation of certain crop types (e.g., sorghum) generally leads to very low SOC contents. (ii) Intensive intercropping (i.e., the simultaneous plantation of different crop types) generally leads to very high SOC rates in the top soil layer (0--20 cm). 

To test hypothesis (i), carbon contents below the 5th percentile of all samples were defined as events in the SOC data set (event time series \texttt{carb}), and the cultivation of sorghum was defined as an event in the crop cover data set (event time series \texttt{crop.sorghum}). In this example, both $\tau$ and $\Delta T$ are again 0, i.e., we are only interested in ``simultaneous'' events. Although the given data sets are no time series, the appropriate data format to choose here is that of event time series (\texttt{ts}) where individual samples take the role of (temporal) observation points. Executing 
\begin{verbatim}
ca.out <- CC.eca.ts(carb,crop.sorghum,...
		... sigtest="poisson")
\end{verbatim}
yields the following results contained in the list \texttt{ca.out}:
\begin{verbatim}
NH precursor: FALSE 
NH trigger: FALSE
p-value precursor: 0.03824319 
p-value trigger: 0.04147892
precursor coincidence rate: 0.2727273
trigger coincidence rate: 0.1875
\end{verbatim}
\noindent
Hence, at the $\alpha=0.05$ significance level, both null hypotheses can be rejected and, thus, a non-random statistical relationship between sorghum plantation and very low SOC values can be deduced. Note that in contrast to the previous example, precursor and trigger coincidence rates are not equal due to the different numbers of events in both data sets. Here, the precursor coindicence rate corresponds to the fraction of plots with low SOC contents on which sorghum was cultivated, whereas the trigger coincidence rate gives the fraction of plots with sorghum plantation on which low SOC contents were observed.


Next, in order to investigate whether intercropping systematically co-occurs with very high top-layer SOC contents (hypothesis (ii)), we define crop covers with at least four different crop types (\texttt{crop.inter}) and SOC contents larger than the 90th percentile (\texttt{carb}) as events and execute
\begin{verbatim}
ca.out <- CC.eca.ts(carb,crop.inter,...
		... sigtest="poisson")
\end{verbatim}
resulting in:
\begin{verbatim}
NH precursor: TRUE
NH trigger: TRUE
p-value precursor: 0.08495326
p-value trigger: 0.07630266
precursor coincidence rate: 0.11
trigger coincidence rate: 0.33
\end{verbatim}
\noindent
In this case, despite the quite high trigger coincidence rate of 0.33, both null hypotheses cannot be rejected at the $\alpha=0.05$ confidence level. The high trigger coincidence rate in this example means that 33\% of the plots characterized by intensive intercropping also show very high SOC values. But since the precursor coincidence rate is rather small, only few plots characterized by high SOC contents have also been cultivated with intercropping. The large difference between trigger and precursor coincidence rate in this example arises again because the numbers of events in the two data sets differ markedly (there are 18 events in the carbon data set and only six events in the crop cover data set). Therefore, to reach a significant $p$-value at $\alpha=0.05$, in the present example at least three of the six intercropping fields would have to coincide with events in the carbon data set (i.e. $r_{p} \ge 0.5$). 

For comparison, we apply the shuffle surrogates-based significance test to the same data set using
\begin{verbatim}
ca.out <- CC.eca.ts(carb,crop.inter,...
		... sigtest="shuffle",reps=10000)
\end{verbatim}
and obtain the following result:
\begin{verbatim}
NH precursor: TRUE
NH trigger: TRUE
p-value precursor: 0.0621
p-value trigger: 0.0522
precursor coincidence rate: 0.11
trigger coincidence rate: 0.33
\end{verbatim}
\noindent
This example illustrates that in cases where the rejection of the null hypothesis is based on $p$-values close to the desired confidence level $\alpha$, the utilization of both significance tests is recommended. Specifically, if we relieve the requested confidence level only slightly (say, $\alpha=0.06$), the null hypothesis of the trigger test could already be rejected, indicating that high top-layer SOC contents are actually supported by intercropping. Although the numbers of ``events'' in both data sets ($N_A=18$ and $N_B=6$) are relatively small in comparison with the sample size ($T=218$), the observed changes in the obtained $p$-values for both precursor and trigger test indicate that the analytical significance test is actually much more conservative than necessary for appropriately testing for the presence of statistical interrelationships between both event series. This suggests that in case of any doubts regarding the validity of the implicit assumptions underlying the analytical significance test, the surrogate-based tests should be preferred.

\section{Conclusions} \label{sec:conclusions}

The new R package \texttt{CoinCalc} allows performing event coincidence analysis (ECA), a novel statistical tool for quantifying the degree of simultaneity between two event series \citep{Donges2011,Donges2015,Rammig2014}, for different types of event series. The package provides six functions: (i) binarization of continuous time series, (ii) data conversion of binary (event) time series to event sequence format, (iii) conversion of event sequences to binary time series, (iv) ECA for event sequences, (v) ECA for binary event time series, and (vi) a plotting function for visualizing events and coincidences. 

Based on two geoscientific example problems, we have illustrated the utilization of the package and interpretation of the obtained results. \texttt{CoinCalc} is freely available via the CRAN repository (\url{www.r-project.org}) and planned to be regularly updated and further extended. At the present stage, \texttt{CoinCalc} provides all necessary functions for performing ECA under relatively general conditions. Future extensions shall include (among others) a more sophisticated surrogate-based significance test for serially dependent event sequences (i.e., series with correlated waiting times between subsequent events) and functions for multivariate and conditional ECA. Moreover, the package will be expanded in order to handle not only pairs of series of vector format, but to also perform ECA between time series of grid points in spatially extended data sets (e.g., with the dimensions latitude, longitude and time), which are typical for climatological data sets or remote sensing products. This extension will also allow analyzing coincidences between time series of different regions, opening the package to a further large field of research questions.
\section*{Acknowledgements}
This work has been financially supported by the German Federal Ministry for Education and Research (BMBF) within the framework of the BMBF Young Investigators Group CoSy-CC$^2$: Complex Systems Approaches to Understanding Causes and Consequences of Past, Present and Future Climate Change (grant no. 01LN1306A). JFS acknowledges funding by the Evangelisches Studienwerk Villigst e.V. The authors are grateful to the SAMPLES project for providing the framework for the soil sample collection, Gustavo Saiz (IMK-IFU, KIT) for supporting the SOC analysis of the soil samples, and Jonathan Donges and Marc Wiedermann for helpful comments on earlier versions of this manuscript and the software package \texttt{CoinCalc} described herein.


 \bibliographystyle{elsarticle-harv}
 \bibliography{Master}

\end{document}